\documentclass[slac_one]{revtex4}
\usepackage{graphicx}
\usepackage{fancyhdr}
\begin{document}

\title{Top Quark Mass in Lepton+Jets Decays at the Tevatron} 

%

\author{C. Schwanenberger (for the CDF and D\O\ collaborations)}
\affiliation{The University of Manchester, Oxford Road, Manchester M13
  9PL, UK}

\begin{abstract}

Various measurements of the top quark mass in the lepton+jets decay channel
of top quark pair production are
presented. The measurements are performed on data samples 
of up to $2.7\ {\rm fb}^{-1}$ of integrated luminosity acquired by the
CDF and D\O\ experiments in Run-II of the Tevatron proton-anti-proton
collider at a center-of-mass energy of $\sqrt{s} =
1.96$~TeV. The new Tevatron combination using up to $2.8\ {\rm fb}^{-1}$ 
of data results in a preliminary world average mass of the top quark of
$m_{\rm top} = 172.4 \pm 1.2$~GeV. This corresponds to a relative
precision of 0.7\%.
\end{abstract}

\maketitle

\thispagestyle{fancy}


\section{Introduction} 
The top quark was discovered in 1995 by
the CDF \cite{CDFdiscovery} and D\O\ \cite{D0discovery} experiments at the Fermilab Tevatron
proton-antiproton collider. The mass of the top quark, which is by far
the heaviest of all quarks, plays an important role in  
electroweak radiative corrections and therefore in constraining the
mass of the Higgs boson. Precise measurements of  
the top quark mass provide a crucial test of the consistency of the
Standard Model (SM) and could indicate a hint of  
physics beyond the SM.   

The Tevatron is still the only place where top quarks can be produced
and studied directly. At the Tevatron, top  
quarks are mostly produced in pairs via the strong interaction. In the
framework of the SM, the top quark decays  
to a $W$ boson and a $b$ quark nearly 100\% of the time. Events from top
quark pair production are classified according 
to the $W$ boson decay channels. 
An event is referred to as ``dilepton'' if both $W$ bosons decay
leptonically, ``all jets'' if both $W$ bosons decay hadronically, 
and ``lepton+jets'' channel if one 
of the $W$ boson decays leptonically in either and electron or muon
and the corresponding neutrino and the other one hadronically.
In this report measurements of the top quark mass in the lepton+jets channel
are presented.

\section{Event Selection}
The selected events are required to 
contain an isolated electron or muon and at
least or exactly four jets with high transverse momenta and large
missing transverse energy.
The dominant background contribution is $W$ boson production with
associated jets ($W$+jets). The second largest background
is 
multijet production where a jet is misidentified as a
lepton and large missing transverse energy is faked. By identifying
$b$ jets in the final state, these background contributions can be
substantially reduced.

\section{Top Quark Mass Measurements}
Different methods to measure the top quark mass are discussed.
``Template Methods''~\cite{CDFtemplate,CDFtemplatealt} have the
advantage of being more straightforward and transparent but
are statistically less accurate. To maximize the
statistical information on the top quark mass extracted from the event
sample, more elaborated but also more complex methods exist as e.g. the
``matrix method''~\cite{CDFmatrix,D0matrix}, the ``ideogram
method''~\cite{D0ideogram} or the ``dynamical likelihood 
method''~\cite{CDFdynlhood}. An alternative method uses the cross
section measurement to extract the top quark
mass~\cite{D0xsec}. Some examples are presented here.

\subsection{Template Method}
Distributions of variables that are strongly correlated
with the top quark mass are derived as templates in Monte Carlo simulations for
different top mass hypotheses. They are compared to the
measured distribution in order to extract the top quark mass from data.

In \cite{CDFtemplate} two variables are used. First, the
reconstructed top quark
mass $m_t^{\rm reco}$ is derived
by minimizing a $\chi^2$-like function to the over-constrained
kinematics of the $t\bar{t}$ system using the measured 4-vectors of the
charged lepton and the jets, the measured missing transverse energy and
$b$-tagging information. It is summed over all possible
assignments of partons to jets. The well-known $W$ mass is used to
constrain the invariant mass of the electron/muon-neutrino pair from
the leptonic $W$ decay and the dijet mass from the hadronic $W$
decay. The top quark and antitop quark masses are constrained to be
equal within the top width. 
As second variable the dijet mass of
the hadronically decaying $W$ boson is used. It constrains the
so-called jet energy scale {\em in situ} which represents the largest
systematic uncertainty in top quark mass measurements. 

The values of both variables for candidate events with at least 1
$b$-tag are compared to a two-dimensional probability density function
derived by applying a kernel density estimation to fully simulated MC
events for different values of the top quark mass and jet energy
scale. The measurement is performed simultaneously in the lepton+jets
and dilepton
channels. The 2-dimensional likelihood is presented in
Fig.~\ref{2Dlhood} (upper left). As a result, a top quark mass of 
$m_{\rm top} = 171.9 \pm 1.7 ({\rm stat}) \pm 1.0 ({\rm syst})
$~GeV was derived. The total accuracy is thus $\pm 1.1$\%.

\subsection{Alternative Template Method}
To be less dependent on the largest systematic uncertainty of the
previous method, in \cite{CDFtemplatealt} quantities are used with
minimal dependence on the jet energy scale. One variable is the
transverse decay length of $b$-tagged jets, and the other is the
transverse momentum of the lepton. Both quantities are roughly
linearly dependent on the top mass. Their combination significantly
reduces the statistical uncertainty because the statistical resolution
in the top mass determination is similar and they are approximately
uncorrelated. As a result, the top quark mass was measured to $m_{\rm top} = 176.7 \pm 6.2 ({\rm stat}) \pm 3.0 ({\rm syst})
$~GeV. This corresponds to a total uncertainty of 3.9\%. Since the
result is statistically limited, it will improve with more data added
in the future, or if the measurement is performed at the Large Hadron
Collider (LHC).

\subsection{Matrix Element Method}
For the matrix element method a probability is calculated for each
event as a function of the assumed top quark mass $m_{\rm top}$ and an
overall multiplicative scale factor JES for jet energies. The factor
JES is fitted {\em in situ} in data, simultaneously with the top quark
mass by using information from the invariant mass of the hadronically
decaying $W$ boson. For every event, this mass is constrained 
to be equal to the known value for the $W$ mass. The probabilities
from all events in the sample are then combined to obtain a
probability as a function of $m_{\rm top}$ and JES, and the top quark
mass is extracted by finding the values that maximize this probability. 

The analyses performed by the CDF~\cite{CDFmatrix} and
D\O~\cite{D0matrix} experiments are very similar. One main difference
is the treatment of background. While in~\cite{CDFmatrix} a neural
network discriminant is used to distinguish signal from background,
in~\cite{D0matrix} the probability for one event is composed from
probabilities not only for top quark pair production signal but also
for $W$+jets background. 

\begin{figure*}[t]
\centering
\includegraphics[width=85mm]{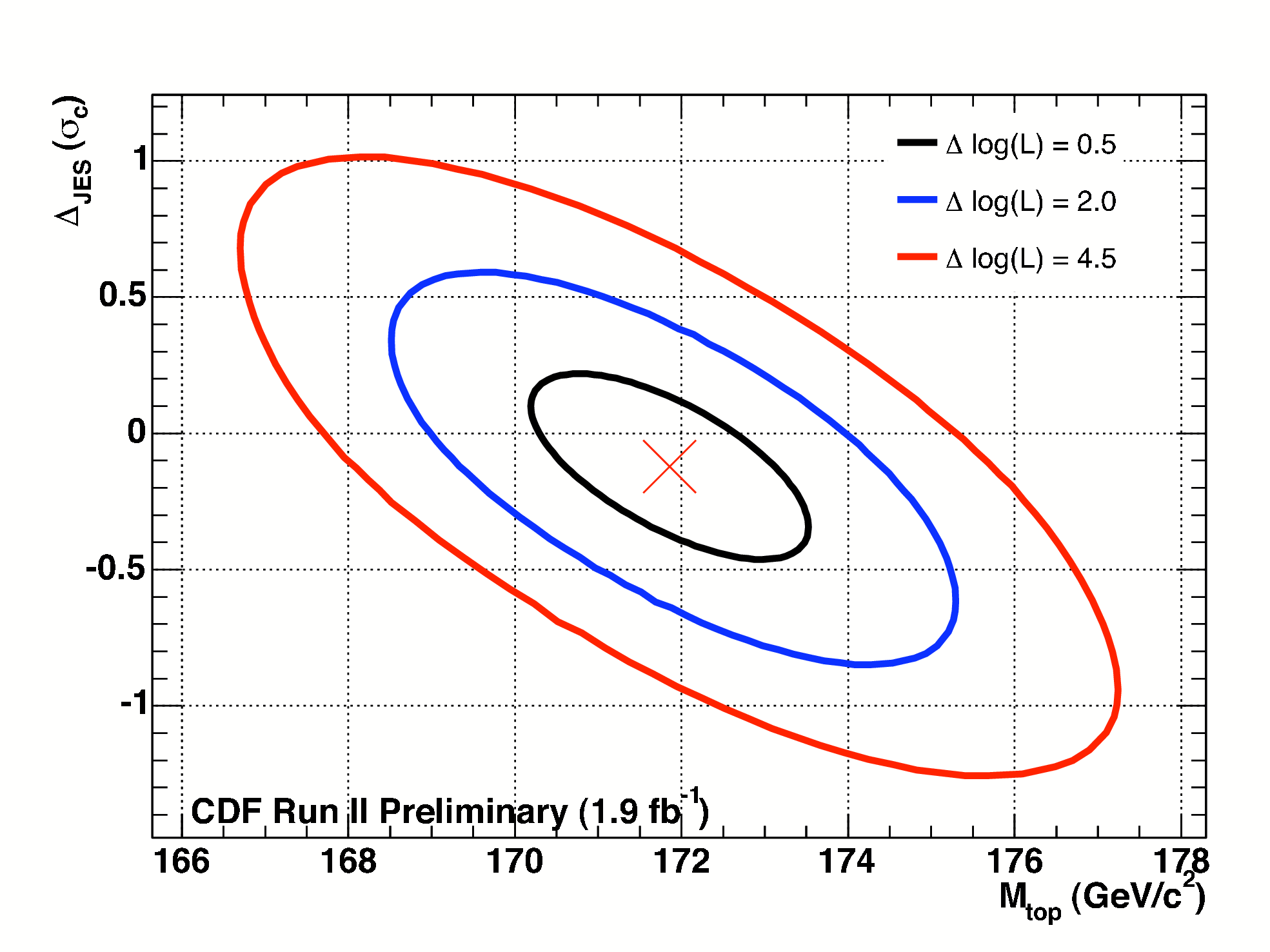}
\includegraphics[width=90.5mm]{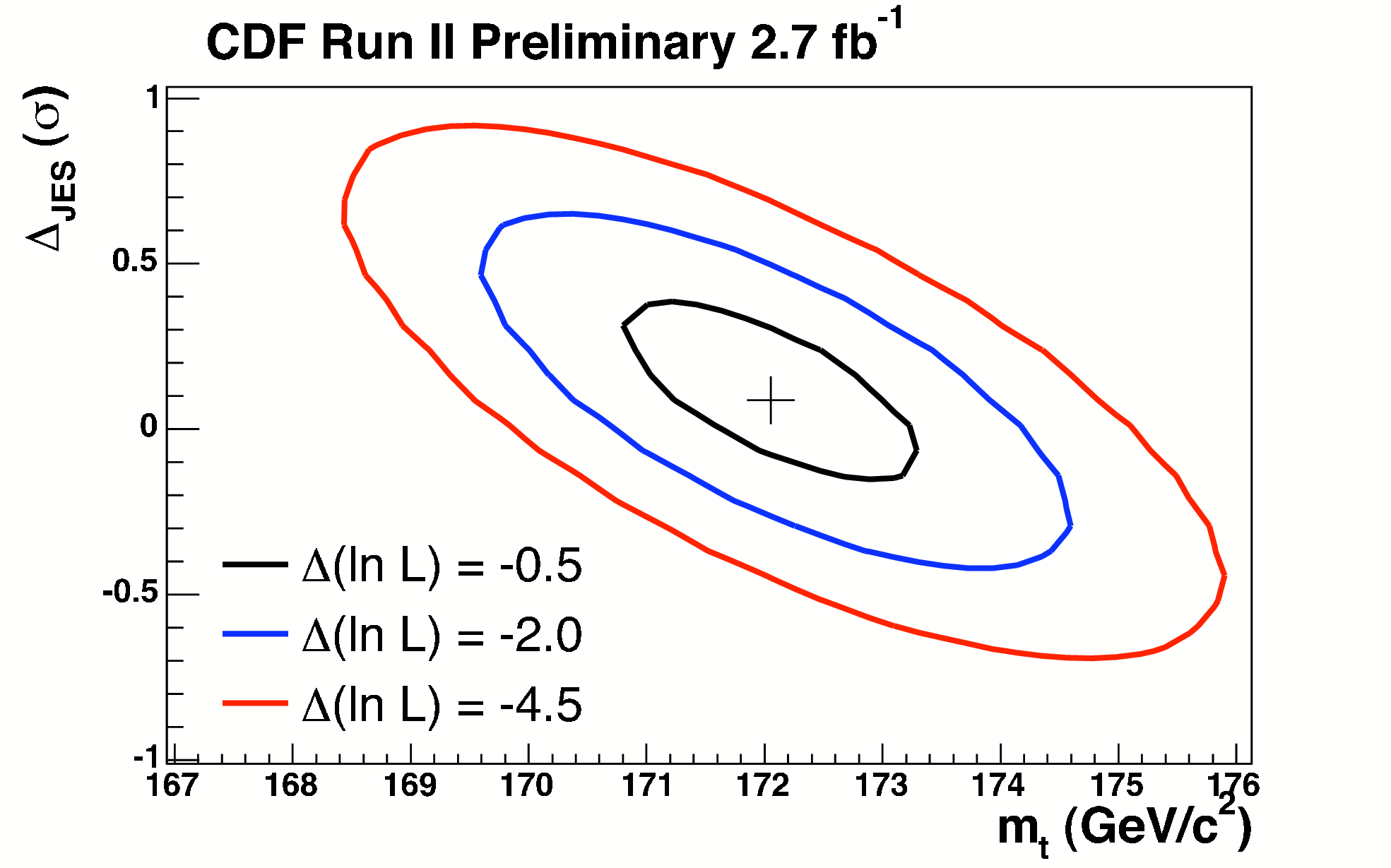}\\
\vspace*{-0.2cm}\hspace{-0.35cm}\includegraphics[width=88mm]{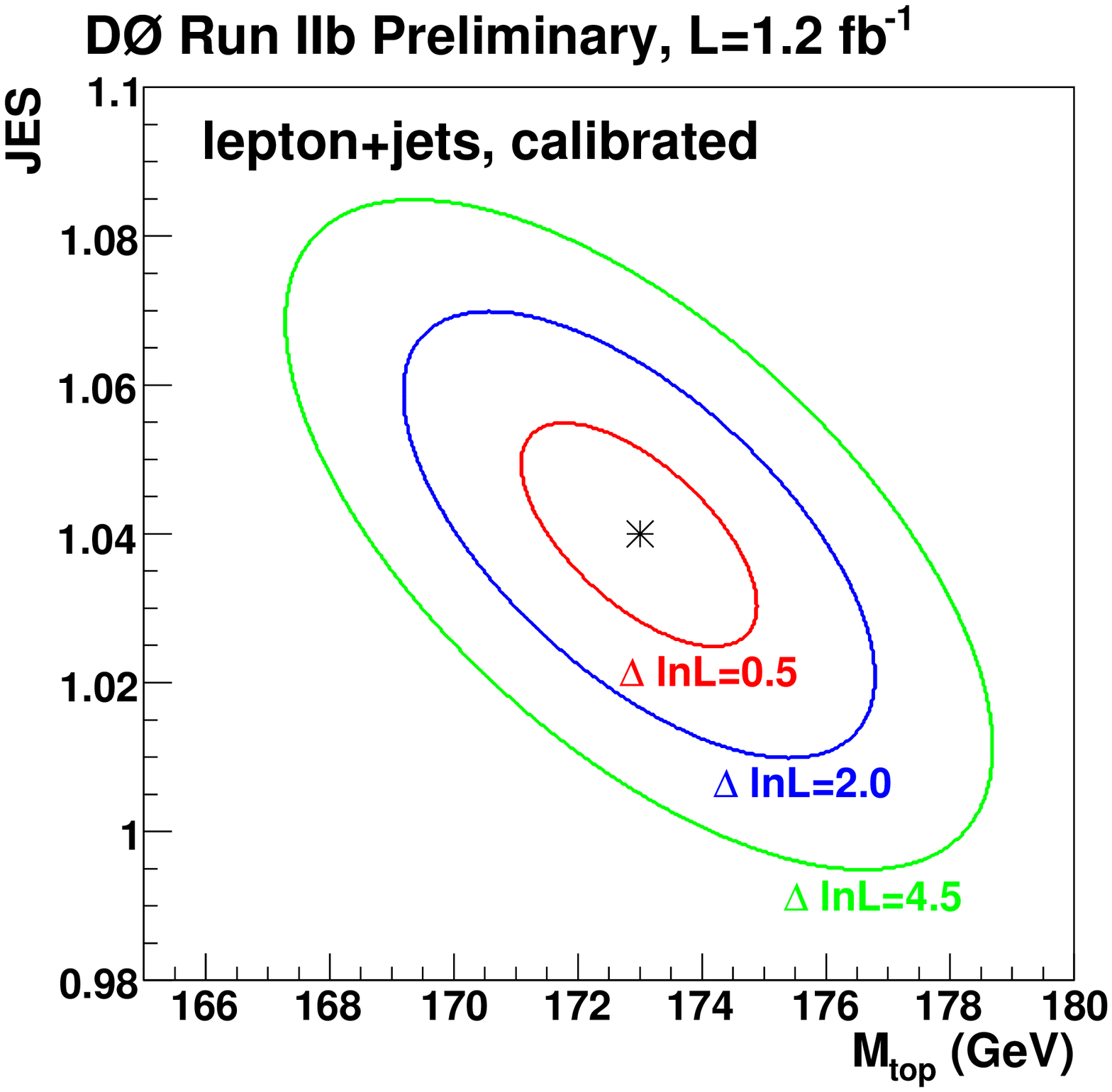}
\hspace{0.15cm}\includegraphics[width=88mm]{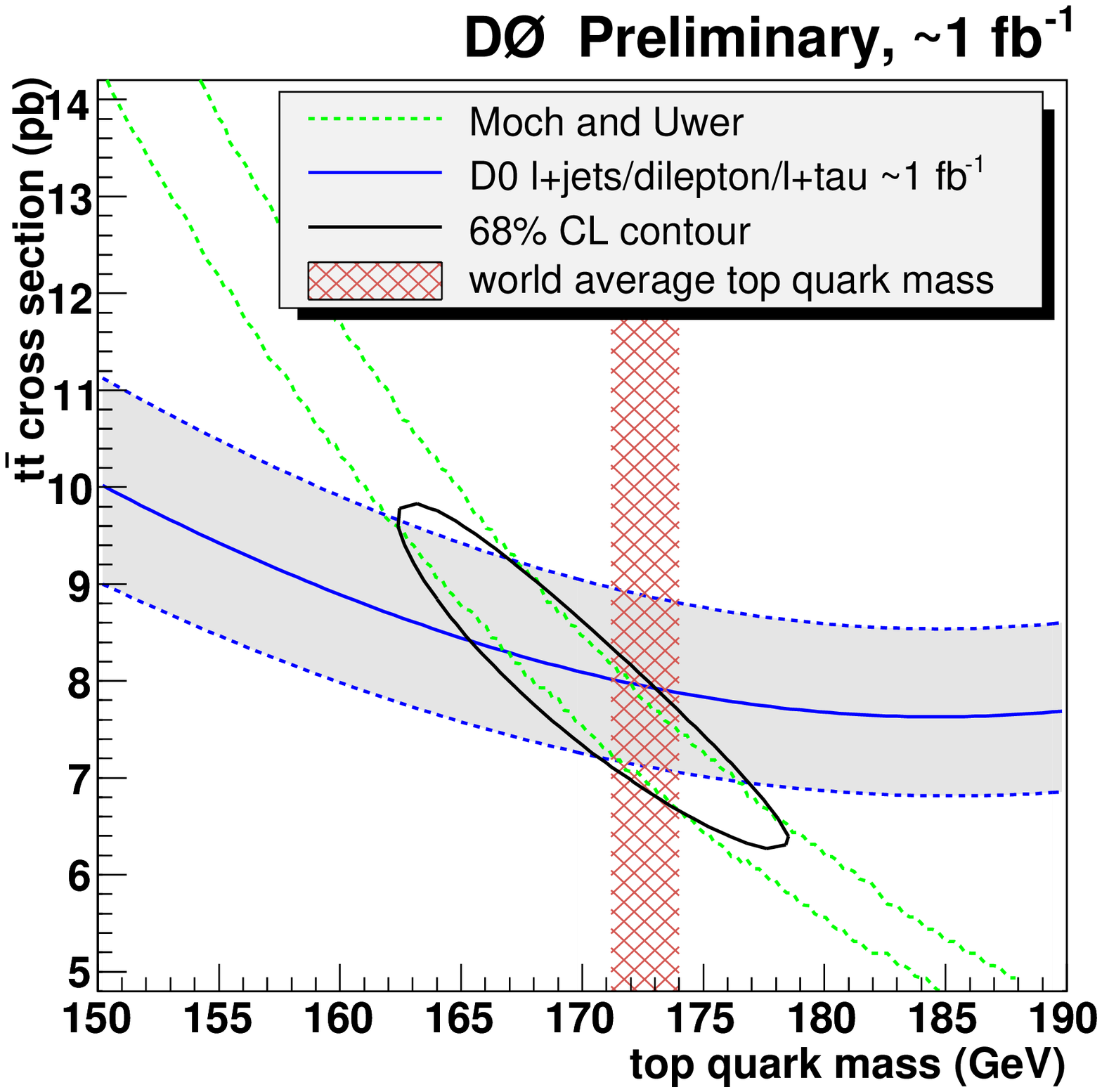}
\caption{2D-likelihood on data events including the $1\sigma$,
  $2\sigma$ and $3\sigma$ uncertainties for the CDF template
  method~\cite{CDFtemplate} (upper left), the CDF matrix element method~\cite{CDFmatrix} (upper right) and the D\O\ matrix element method~\cite{D0matrix} (lower
  left). The latter was derived from a subsample of $1.2\ {\rm 
  fb}^{-1}$ while the final result as given in the text was derived
  from a $2.2\ {\rm fb}^{-1}$ data set. The lower right plot shows
  the top pair production cross section as a function of the top quark
  mass for the combination of the
  lepton+jets, dilepton and tau+lepton channels~\cite{peters}, the
  theory prediction 
  from~\cite{moch} and the combined $1\sigma$ likelihood.
} \label{2Dlhood}
\end{figure*}

For both measurements pseudo-experiments using a large  pool of fully
simulated MC events are performed to
calibrate the method, correcting for biases to ensure 
that the fitted parameters represent true values and that the
estimated errors can be trusted. The two-dimensional
likelihoods on data events are shown in Fig.~\ref{2Dlhood} (upper
right for CDF and lower left for D\O). As a
result the top quark mass is measured to $m_{\rm top} = 172.2 \pm 1.0
({\rm stat}) \pm 1.3 ({\rm syst})$~GeV by CDF~\cite{CDFmatrix} and
$m_{\rm top} = 172.2 \pm 1.0 
({\rm stat}) \pm 1.4 ({\rm syst})$~GeV by D\O~\cite{D0matrix}. The
total uncertainties for both measurements are $\pm1.0$\%. They are
systematically limited. The largest sources of systematic
uncertainties apart from the simultaneous inclusion of JES are given
by residual JES, in particular of $b$-jets, and theoretical
uncertainties in signal and background modeling. Currently both
experiments are undertaking large efforts to get a uniform treatment
of all uncertainties where ever possible.

\subsection{Cross Section Method}
The value of a quark mass is renormalization-scheme dependent. Thus it
is important to extract this parameter using a well-defined
renormalization scheme. Direct top quark mass measurements as
discussed previously compare
measured distributions to distributions simulated by leading-order MC
generators as a function of the top quark mass. The input mass for
these MC generators is not in a well-defined renormalization scheme
leading to an uncertainty in its definition. This can be avoided
when the top
quark mass is extracted by comparing top pair production cross section
measurements to fully  
inclusive theoretical calculations in higher-order QCD. They include soft
gluon resummations and represent the most  
complete calculations available. Furthermore, they are worked out
using the pole mass 
definition for the top quark which is thus the parameter extracted
here.  

The experimental and theoretical cross sections
as a function of the
top mass were fitted using third-order polynomials in the top quark
mass and Likelihoods were defined. The theoretical and experimental
likelihoods are multiplied to obtain a joint
likelihood as a function of the top mass. The extracted top mass is
given by the minimum of the likelihood function. As an example, the
combination of the $t\bar{t}$ cross sections derived in the
lepton+jets~\cite{Rb}, 
dilepton~\cite{dilepton} and lepton+tau~\cite{taulep} decay channels as
described in~\cite{peters} is compared to a next-to-leading order
(NLO) QCD calculation that includes all
next-to-next-to-leading logarithms (NNLL) that are relevant in
next-to-next-to-leading order (NNLO) QCD~\cite{moch}. This is shown
together with the joint likelihood in
Fig.~\ref{2Dlhood} (lower right). The extracted top
quark pole mass is $M_{\rm top} = 169.6\ ^{+5.4}_{-5.5}$~GeV. The
total uncertainty is thus $\pm 3.2$\%.

\section{Top Quark Mass Combinations}
Taking correlated uncertainties properly into account the CDF and
D\O\ collaborations have derived separate combinations 
of published Run-I (1992--1996) measurements with the most recent 
preliminary Run-II (2001--present) measurements using up to $2.8\ {\rm
  fb}^{-1}$ of data. The new CDF combination~\cite{CDFcombi} gives
$m_{\rm top} = 172.4 \pm 1.0  
({\rm stat}) \pm 1.3 ({\rm syst})$~GeV, the new D\O\
combination~\cite{D0combi} $m_{\rm top} = 172.8 \pm 1.0 
({\rm stat}) \pm 1.3 ({\rm syst})$~GeV. These results were combined to
the new world average~\cite{worldave} of $m_{\rm top} = 172.4 \pm 0.7  
({\rm stat}) \pm 1.0 ({\rm syst})$~GeV. This corresponds to a relative
precision of 0.7\% on the top quark mass.

\section{Conclusions}
Various measurements of the top quark mass in the lepton+jets channel
and combinations with other channels were presented. The most
precise measurements today are systematically limited. The main challenge
right now is thus to work on reducing the experimental and theoretical
uncertainties where ever possible. 
To get a better understanding of the
interpretation of the measured quantity with respect to the applied
renormalization scheme will be an important task for the future.


\end{document}